\documentclass[showpacs,superscriptaddress,floatfix,twocolumn]{revtex4}
\usepackage{graphicx}
\usepackage{amsmath}
\usepackage{amssymb}
\usepackage{tabularx}
\usepackage{multirow}
\newcommand{\PreserveBackslash}[1]{\let\temp=\\#1\let\\=\temp}

\begin{document} 

\setlength{\textheight}{1.007\textheight}

\title{Diffusion of transcription factors can drastically enhance the noise in gene expression} 
 
\author{Jeroen S. van Zon} 
\affiliation{Division of Physics and Astronomy, Vrije Universiteit, De 
Boelelaan 1081, 1081 HV Amsterdam, The Netherlands.} 
 
\author{Marco J. Morelli} 
\affiliation{FOM Institute for Atomic and Molecular Physics, 
Kruislaan 407, 1098 SJ Amsterdam, The Netherlands.} 
 
\author{Sorin T\u{a}nase-Nicola} 
\affiliation{FOM Institute for Atomic and Molecular Physics, 
Kruislaan 407, 1098 SJ Amsterdam, The Netherlands.} 
 
\author{Pieter Rein ten Wolde} 
\affiliation{FOM Institute for Atomic and Molecular Physics, 
Kruislaan 407, 1098 SJ Amsterdam, The Netherlands.} 
 
\date{\today}

\begin{abstract}  
 
We study by simulation the effect of the diffusive motion of repressor 
molecules on the noise in mRNA and protein levels in the case of a 
repressed gene. We find that spatial fluctuations due to diffusion 
can drastically enhance the noise in gene expression. For a fixed 
repressor strength, the noise due to diffusion can be minimized by 
increasing the number of repressors or by decreasing the rate of the 
open complex formation.  We also show that the effect of spatial 
fluctuations can be well described by a two-step kinetic scheme, where 
formation of an encounter complex by diffusion and the subsequent 
association reaction are treated separately. Our results also 
emphasize that power spectra are a highly useful tool for studying the 
propagation of noise through the different stages of gene expression.\\ 
 
\noindent key words: gene expression, noise, systems biology, computer simulation 
\end{abstract} 
 
 

\maketitle

\noindent  
\section{Introduction} 
 Cells process information from the outside and regulate their internal 
state by means of proteins and DNA that chemically and physically 
interact with one another. These biochemical networks are often highly 
stochastic, because in living cells the reactants often occur in small 
numbers. This is particularly important in gene 
expression~\cite{McAdams97,Elowitz02,Ozbudak02}, where transcription 
factors are frequently present in copy numbers as low as tens of 
molecules per cell. While it is generally believed that biochemical 
noise can be detrimental to cell function~\cite{Rao02}, it is 
increasingly becoming recognized that noise can also be beneficial to 
the organism~\cite{Kaern05}. Understanding noise in gene expression is 
thus important for understanding cell function, and this observation 
has recently stimulated much theoretical and experimental work in this 
direction~\cite{Rao02,Kaern05}. However, the theoretical analyses 
usually employ the zero-dimensional chemical master 
equation~\cite{kampenbook,gillespie76}. This approach takes into 
account the discrete character of the reactants and the probabilistic 
nature of chemical reactions. It does assume, however, that the cell 
is a `well-stirred' reactor, in which the particles are uniformly 
distributed in space at all times; the reaction rates only depend upon 
the global concentrations of the reactants and not upon the spatial 
positions of the reactant molecules. Yet, in order to react, reactants 
first have to move towards one another. They do so by diffusion, or, 
in the case of eukaryotes, by a combination of diffusion and active 
transport. Both processes are stochastic in nature and this could 
contribute to the noise in the network. Here, we study by computer 
simulation the expression of a single gene that is under the control 
of a repressor $R$ in a spatially-resolved model. We find that at low 
repressor concentration, {\em i.e.} $[R] < 50 {\rm nM}$, the noise in 
gene expression is dominated by the noise arising from the diffusive 
motion of the repressor molecules. Our results thus show that spatial 
fluctuations of the reactants can be an important source of noise in 
biochemical networks. 
 
Our analysis reveals that in gene expression significant fluctuations 
occur on both small and large length and time scales. As expected from 
earlier work~\cite{Swain02,Paulsson04,Elowitz05}, the fluctuations on 
long time scales are predominantly due to protein degradation; we 
assume that proteins are degraded by dilution, which means that the 
relaxation rate of this process is on the order of an hour. Our 
results, however, also elucidate an important process on much smaller 
length and time scales. It is associated with the competition between 
repressor and RNA polymerase (RNAP) for binding to the promoter. When 
a repressor molecule dissociates from the DNA, it can rebind very 
rapidly, {\em i.e.} on a time scale of microseconds, or less. This 
rebinding time is so short that when a repressor molecule has just 
dissociated, the probability that a RNAP will bind before the 
repressor molecule rebinds, is very small. As a result, a repressor 
molecule will on average rebind many times, before it eventually 
diffuses away from the promoter and a RNAP molecule, or another 
repressor molecule, can bind to the promoter. This process of rapid 
rebindings decreases the effective dissociation rate, and this 
increases the noise in gene expression. 
 
Clearly, fluctuations in gene expression span orders of magnitude in 
length and time scales. This means that the simulation technique 
should be sufficiently detailed to resolve the events at small length 
and time scales, yet also efficient enough to access the long length 
and time scales. Recently, several simulation techniques have been 
developed for the stochastic modeling of reaction-diffusion 
systems~\cite{Elf04,Ander04}. These techniques, however, do not 
satisfy both criteria: they either describe the system in a 
coarse-grained way, {\em i.e.} on the level of local concentrations 
rather than single particles~\cite{Elf04,Ander04}, or are too slow to 
accurately model the dynamics on the long time 
scales~\cite{Andrews04}. Our simulations have been made possible via 
the use of our recently developed Green's Function Reaction Dynamics 
(GFRD) algorithm~\cite{VanZon05,VanZon05_2}. GFRD is an event driven 
algorithm that uses Green's functions to combine in one step the 
propagation of the particles in space with the reactions between 
them. The event-driven nature of the algorithm makes it particularly 
useful for problems, such as gene expression, in which the events are 
distributed over a wide range of length and time scales: the algorithm 
takes small steps when the reactants are close to each other -- such 
as when a repressor molecule has just dissociated from the DNA -- 
while it takes large jumps in time and space when the molecules are 
far apart from each other -- like when the repressor molecule has 
eventually diffused away from the promoter. The event-driven nature of 
GFRD makes it orders of magnitude more efficient than brute-force 
particle-based algorithms~\cite{VanZon05_2} and this has allowed us to 
simulate gene expression on the relevant biological time scales of 
hours. 
 
Several publications \cite{ko91, kepler01, karmakar04, pirone04, 
simpson04, bialek05, hornos05} have discussed the effect of 
fluctuations in the binding of transcription factors to their site on 
the DNA (called operator) on the noise in gene expression. Most of  
these models are relatively simple, ignoring, for instance, production 
of mRNA \cite{kepler01, karmakar04, pirone04, hornos05}. Moreover all 
these studies, with the exception of \cite{metzler01,bialek05}, ignore 
the role of the spatial fluctuations of the transcription factors. Our 
aim is to study gene expression in a biologically meaningful model. We 
have therefore constructed a rather detailed model, although we will 
also use minimal models that can be studied analytically, in order to 
interpret the simulation results. The full model, which is described 
in the next section, contains the diffusive motion of repressor 
molecules, open complex formation, promoter clearance, transcription 
elongation and translation~\cite{Record96}.  
 
 In section~\ref{sec:simres}, we discuss the simulation results for 
both the noise in mRNA and protein level. The results reveal that for 
$[R]<50 {\rm nM}$, the noise in the spatially-resolved model can be 
more than five times larger than the noise in the well-stirred model. 
We also show that a cell could minimize the effect of spatial 
fluctuations, either by tuning the open complex formation rate or by 
changing the number of repressors and their affinity for the binding 
site on the DNA. In section~\ref{sec:operbind}, we elucidate the 
origin of the enhanced noise in the spatially resolved model. In the 
subsequent section, we show that in the model employed here the effect 
of spatial fluctuations can be quantitatively described by a 
well-stirred model in which the reaction rates for repressor binding 
and unbinding are appropriately renormalized; however, as we discuss 
in the last section, we expect that in a more refined model the effect 
of diffusion will be more complex, impeding such a simplified 
description. In section~\ref{sec:ps}, we discuss how the operator 
state fluctuations propagate through the different stages of gene 
expression using power spectra for the operator state, elongation 
complex, mRNA and protein. The results show that these power spectra 
are highly useful for unraveling the dynamics of gene expression. We 
hope that this stimulates experimentalists to measure power spectra of 
not only mRNA and protein levels~\cite{Austin06}, but also of the 
dynamics of transcription initiation and elongation using {\em e.g.} 
magnetic tweezers~\cite{Revyakin04}. As we argue in the last section, 
such experiments should make it possible to determine the importance 
of spatial fluctuations for the noise in gene expression.\section{Model} 
\subsection{Diffusive motion of repressors} 
\label{sec:diffusion} 
We explicitly simulate the diffusive motion of the repressor 
molecules in space. However, since the experiments of Riggs {\em et 
al.}~\cite{Riggs70} and the theoretical work of Berg, Winter, and Von 
Hippel~\cite{Berg81}, it is well known that proteins could 
find their target sites via a combination of 1D sliding along the DNA 
and 3D diffusion through the cytoplasm -- ``hopping'' or ``jumping'' from one 
site on the DNA to another. This mechanism could speed up the search 
process and make it faster than the rate at which 
particles find their target by free 3D diffusion; this rate is given 
by $k = 4 \pi \sigma D_3 [R]$, where $\sigma$ is the cross section, 
which is on the order of a protein diameter or DNA diameter, $D_3$ is 
the diffusion constant of the protein in the cytoplasm, and $[R]$ is 
the concentration of the (repressor) protein. However, while it is 
clear that the mechanism of 3D diffusion and 1D sliding could 
potentially speed up the search process, whether this mechanism in 
living cells indeed drastically reduces the search time is still under 
debate~\cite{Halford04}. In this context, it is instructive to discuss 
 the two main results of recent studies on this 
topic~\cite{Gerland02,Coppey04,Slutsky04,Halford04,Klenin06,Hu06}. The first is 
that the mean search time $\tau$ is given by~\cite{Hu06} 
\begin{equation} 
\label{eq:tau} 
\tau \sim \frac{L}{\lambda} \left[\frac{\lambda^2}{D_1} + 
\frac{r^2}{D_3}\right], 
\end{equation} 
where $L$ is the total length of the DNA, $\lambda$ is the average 
distance over which the protein slides along the DNA before it 
dissociates, $D_1$ is the diffusion constant for sliding, $r$ is 
the typical mesh size in the nucleoid, and $D_3$ is the diffusion 
constant in the cytoplasm. This formula has a clear 
interpretation~\cite{Hu06}: $\lambda^2/D_1$ is the sliding time, 
$r^2/D_3$ is the time spent on 3D diffusion, the sum of these terms is 
thus the time to perform one round of sliding and diffusion, and 
$L/\lambda$ is the total number of rounds needed to find the 
target. The other principal result is that the search time is minimized 
when the sliding distance $\lambda$ is 
\begin{equation} 
\label{eq:lambda} 
\lambda = \sqrt{\frac{D_1}{D_3}} r. 
\end{equation} 
Under these conditions, a protein spends equal amounts of time on 3D 
diffusion and 1D sliding (a protein is thus half the time bound to the 
DNA). Eq.~\ref{eq:lambda} is a useful result, because it shows that 
the average sliding distance $\lambda$ depends upon the ratio of 
diffusion constants and on the typical mesh size in the nucleoid. If 
we now assume that $D_1$ and $D_3$ are equal (which is not obvious 
given that proteins bind relatively strongly to DNA -- $D_1$ could 
thus very well be much smaller than $D_3$) and if we take the mesh 
size to be given by $r\sim \sqrt{v/L}$~\cite{Hu06}, where $v \approx 1 \mu {\rm 
m}^3$ is the volume of an {\em Escherichia coli} cell and $L \approx 10^3 \mu 
{\rm m}$, we find that $\lambda \approx 10 {\rm nm}$ (30 bp). This 
corresponds to the typical diameter of a protein or DNA double helix 
and is thus not very large. Interestingly, recent experiments seem to 
confirm this: experiments from Halford {\em et al.} on restriction 
enzymes (EcoRV and BbcCI) with a series of DNA substrates with two 
target sites and varying lengths of DNA between the two sites, 
suggest that under the {\em in vivo} conditions, sliding is indeed 
limited to relatively short distances, {\em i.e.} to distances less 
than 50 bp ($\approx 16 {\rm nm}$)~\cite{Stanford00,Gowers05}. 
 
Now, it should be realized that on length scales beyond the sliding 
length, the motion is essentially 3D diffusion: the sliding/hopping 
mechanism corresponds to 3D diffusion with a jump distance given by 
the sliding distance~\cite{Gerland02}. Moreover, since the sliding 
distance is only on the order of a particle diameter, as discussed 
above, we have therefore decided to model the motion of the repressor 
molecules as 3D diffusion. But it should be remembered that on length 
scales smaller than $10 - 30 {\rm nm}$, this approach is not 
correct. We discuss the implications of this for our results in the 
Discussion section. 
\subsection{Transcription and Translation} 
\label{sec:TT} 
Most repressors bind to a site that (partially) overlaps with the core 
promoter -- the binding site of the RNA polymerase (RNAP). When a 
repressor molecule is bound to its operator site, it prevents RNAP 
from binding to the promoter, thereby switching off gene 
expression. Only in the absence of a repressor on the operator site, 
can RNAP bind to the promoter and initiate transcription and translation, ultimately resulting 
in the production of a protein. We model this by the following 
reaction network: 
 \begin{eqnarray} 
\label{eq:gene_exp1} 
O + R &\overset{k_{\rm fR}}{\underset{k_{\rm bR}}{\rightleftarrows}}& OR \\ 
\label{eq:gene_exp2} 
O &\overset{k_{\rm fRp}}{\underset{k_{\rm bRp}}{\rightleftarrows}}& ORp \\ 
\label{eq:gene_exp3} 
ORp &\underset{k_{\text{OC}}}{\rightarrow}& ORp^* \\ 
\label{eq:gene_exp4} 
ORp^* &\underset{t_{\text{clear}}}{\rightarrow}& T + O \\ 
\label{eq:gene_exp5} 
T &\underset{t_{\text{elon}}}{\rightarrow}& M \\ 
\label{eq:gene_exp6} 
M &\underset{k_{\text{dm}}}{\rightarrow}& \varnothing \\ 
\label{eq:gene_exp7} 
M &\underset{k_{\text{ribo}}}{\rightarrow}& M + M_{\rm ribo}\\ 
\label{eq:gene_exp8} 
M_{\rm ribo} &\underset{t_{\text{trans}}}{\rightarrow}& P \\ 
\label{eq:gene_exp9} 
P &\underset{k_{\text{dp}}}{\rightarrow}& \varnothing 
\end{eqnarray} 
Eqs. \ref{eq:gene_exp1} and \ref{eq:gene_exp2} describe the 
competition between the binding of the repressor $R$ and the RNAP 
molecules $Rp$ to the promoter ($O$ is the operator site). In our 
simulation we fix the binding site $O$ in the center of a container 
with volume $V=1\mu {\rm m}^3$, comparable to the volume of a single 
{\em E. coli} cell.  We simulate both the operator site $O$ 
and the repressor molecules as spherical particles with diameter 
$\sigma = 10$nm. The operator site $O$ is surrounded by $N_R$ 
repressor molecules that move by free 3D diffusion (see previous 
section) with an effective diffusion constant $D= 1 \mu {\rm m}^2 s^{-1}$, as 
has been reported for proteins of a similar size \cite{elowitz99}. 
The intrinsic forward rate $k_{\text{fR}} = 6\cdot10^9 M^{-1} s^{-1}$ 
for the repressor particles $R$ at contact is estimated from the 
Maxwell-Boltzmann distribution \cite{VanZon05}. The backward rate 
$k_{\text{bR}}$ depends on the interaction between the DNA binding 
site of the repressor and the operator site on the DNA and varies 
greatly between different operons, with stronger repressors having a 
lower $k_{\text{bR}}$. In our simulations, we vary $k_\text{bR}$ 
between $1 - 0.01\ s^{-1}$, as discussed in more detail below. The 
concentration of RNAP is much higher than that of the 
repressor~\cite{Bremer03}. Because of this we treat the RNAP as distributed 
homogeneously within the cell and we do not to take diffusion of RNAP 
into account explicitly. Instead, RNAP associates with the promoter 
with a diffusion-limited rate $k_\text{fRp} = 4 \pi \sigma D [Rp]$. In 
our simulations, the concentration of free RNAP is $[Rp] = 0.5 \mu M$ 
\cite{Bremer03}, leading to a forward rate $k_{\text{fRp}}=38 
s^{-1}$. Finally, the backward rate $k_{\text{bRp}}=0.5$ is determined 
such that $K_{\text{eq}} = 4 \pi \sigma D / k_{\text{bRp}} = 1.4 \cdot 
10^9 M^{-1}$ \cite {McClure83}. 
 
Transcription initiation is described by Eqs. \ref{eq:gene_exp3} and 
\ref{eq:gene_exp4}. Before productive synthesis of RNA occurs, first 
the RNAP in the RNAP-promoter complex $ORp$ unwinds approximately one 
turn of the promoter DNA to form the open complex $ORp^*$.  The open 
complex formation rate $k_\text{OC}$ has been measured to be on the 
order of $0.3 - 3 s^{-1}$ \cite{Revyakin04}. We approximate open 
complex formation as an irreversible reaction. Some experiments find 
this step to be weakly reversible \cite{Revyakin04}. However, adding a 
backward reaction to the model did not change the dynamics of the 
system in a qualitative way, as long as the backward rate is smaller 
than $k_\text{OC}$, which is in agreement with experimental 
results. After open complex formation, RNAP must first escape the 
promoter region before another RNAP or repressor can bind. Since 
elongation occurs at a rate of $50-100$ nucleotides per second and 
between $30-60$ nucleotides must be cleared by RNAP before the 
promoter is accessible, a waiting time of $t_\text{clear} = 1 s$ is 
required before another binding can occur. Since promoter clearance 
consists of many individual elongation events that obey Poisson 
statistics individually, we model the step as one with a fixed time delay 
$t_\text{clear}$, not as a Poisson process with rate 
$1/t_\text{clear}$. 
 
Eqs. \ref{eq:gene_exp5}-\ref{eq:gene_exp9} describe the dynamics of 
mRNA and protein numbers. After clearing the promoter region, RNAP 
starts elongation of the transcript $T$. As for clearance, the 
elongation step is modeled as a process with a fixed time delay 
$t_\text{elon} = 30 s$, corresponding to an elongation rate of 
$50-100$ nucleotides per second and a $1500$ bp gene. When a mRNA $M$ 
is formed, it can degrade with a rate $k_\text{dm}$. Here, the mRNA 
degradation rate is determined by fixing the average mRNA 
concentration in the unrepressed state, as described below. 
Furthermore, a mRNA molecule can form a mRNA-ribosome complex $M_{\rm 
ribo}$ and start translation. We assume that $b=5$ proteins are 
produced on average from a single mRNA molecule~\cite{Ozbudak02}, so 
that the start of translation occurs at a rate $k_\text{ribo} = b\ 
k_\text{dm}$. After a fixed time delay $t_\text{trans}=30 s$ a protein 
$P$ is produced. The mRNA is available for ribosome binding 
immediately after the start of translation. Due to the delay in 
protein production, $M$ can start to be degraded, while the 
mRNA-ribosome complex $M_{\rm ribo}$ is still present; $M$ thus 
represents the mRNA leader region rather than the entire mRNA 
molecule. Finally, the protein $P$ degrades at a rate $k_\text{dp}$, 
which is determined by the requirement that the average protein 
concentration in the unrepressed state has a desired value, as we 
describe now.

We vary the free parameters in the reaction network described in 
Eqs. \ref{eq:gene_exp1}-\ref{eq:gene_exp9} --  $N_R$, $k_{\rm bR}$, 
$k_{\rm dm}$, $k_{\rm dp}$ -- in the following way: first, we choose the 
concentration of mRNA and protein in the absence of repressor 
molecules.  In this case, tuning of the concentrations is most 
straightforward by adjustment of the mRNA and protein decay rates 
$k_\text{dm}$ and $k_\text{dp}$. For the above reaction network one 
can show that the average mRNA number $N_M$and protein number $N_P$ is 
given by 
\begin{eqnarray} 
\label{eq:rate_eq1} 
N_M &=& \frac{ K_4 K_1 V }{K_2 N_R + V ( 1 + K_1 ( 1 + K_3 ) )}, \\ 
\label{eq:rate_eq2} 
N_P &=& K_5 N_M, 
\end{eqnarray}   
where $K_1 = k_\text{fRp}/(k_{bRp}+k_\text{OC})$, $K_2 = 
k_\text{fR}/k_\text{bR}$, $K_3 = k_\text{OC} t_\text{clear}$, $K_4 = 
k_\text{OC} / k_\text{dm}$ and $K_5 = k_\text{ribo}/k_\text{dp}$ are 
equilibrium constants, $V$ is the volume of the cell and $N_R$ is the 
total number of repressors. The unrepressed state corresponds to 
$N_R=0$. In our simulations, we fix the mRNA and protein numbers in 
the unrepressed state at $N_M = 50$ and $N_P = 2\cdot10^5$. The mRNA 
and protein decay rates then follow straightforwardly from Eqs. 
\ref{eq:rate_eq1} and \ref{eq:rate_eq2}: the mRNA degradation rate is 
$k_{\rm dm} = 0.019 {\rm s}^{-1}$~\cite{Kushner96} and the protein 
degradation rate is $k_{\rm dp} = 2.4 \times 10^{-4} {\rm s}^{-1}$; 
the latter corresponds to protein degradation by dilution with a cell 
cycle time of around 1h. 
 
Next, we determine by what factor these concentrations should decrease in the repressed state. This can be done by changing the number 
of repressors $N_R$ and the repressor backward rate $k_\text{bR}$. We define the repression level $f$ as the transcription initiation 
rate in the absence of repressors, divided by the initiation rate in the repressed state \cite{vilar03}. For a repression level $f$, the 
concentration of mRNA and proteins in the repressed state is a fraction $1/f$ of the concentration in the unrepressed state and it  
follows that 
\begin{equation} 
\label{eq:repression_factor} 
\frac{N_R}{k_\text{bR}} = (f-1) \frac{V( 1+K_1(1+K_3))}{k_\text{fR}}. 
\end{equation} 
Thus, a fixed repression level $f$ does not specify a unique 
combination of $N_R$ and $k_\text{bR}$: increasing the number of 
repressors twofold, while also increasing the repressor backward rate 
by the same factor, gives the same repression level. This means that 
the cell can control mRNA and protein levels in the repressed state 
either by having a large number of repressors that stay on the DNA for 
a short time or by having a small number of repressors, possibly even 
one, that stay on the DNA for a long time. Even though it is 
conceivable that the latter is preferable for economic reasons, there 
is no difference between the two extremes in terms of the average gene 
expression. In our simulations, we vary $N_R$ and $k_{\rm bR}$, but use 
a fixed repression level $f=100$. Consequently, in the repressed 
state, on average $N_\text{M} = 0.5$ and $N_\text{P}=200$. 
\section{Simulation Technique} 
We simulate the above reaction network using Green's Function Reaction 
Dynamics (GFRD) \cite{VanZon05,VanZon05_2}. As discussed above, only 
the operator site $O$ and the repressor particles $R$ are simulated in 
space. All other reactions are assumed to occur homogeneously within 
the cell and are simulated according to the well-stirred model 
\cite{Gillespie77} or with fixed time delays for reaction steps 
involving elongation.  A few modifications with respect to the 
algorithm described in \cite{VanZon05,VanZon05_2} are implemented to 
improve simulation speed. First, we neglect excluded volume 
interactions between repressor particles mutually, as the 
concentration of repressor is very low. This means that the only 
potential reaction pairs we consider are operator-repressor 
pairs. Secondly, we use periodic boundary conditions instead of a 
reflecting boundary, which leads to a larger average time step. As the 
operator site $O$ is both small compared to the volume of the cell and 
is far removed from the cell boundary, this has no effect on the 
dynamics of the system. Finally, as the repressor backward rate 
$k_\text{bR}$ is rather small, the operator site can be occupied by a 
repressor for a time long compared to the average simulation time 
step. If the repressor is bound to the operator site longer than a 
time $L^2/6 D$, where $L$ is the length of the sides of our container, 
the other repressor molecules diffuse on average from one side of the 
box to the other. Consequently, when the repressor eventually 
dissociates from the operator site, the other repressor molecules have 
lost all memory of their positions at the time of repressor 
binding. Here, when a repressor will dissociate after a time longer 
than $L^2/6 D$, we do not propagate the other repressors with GFRD, 
but we only update the master equation and fixed delay reactions. We 
update the positions of the free repressors at the moment that the 
operator site becomes accessible again, by assigning each free 
repressor molecule a random position in the container; the dissociated 
repressor is put at contact with the operator site. We see no 
noticeable difference between this scheme and results obtained by the 
full GFRD algorithm described in Refs. \cite{VanZon05,VanZon05_2}. 
\section{Simulation results: dynamics and noise} 
\label{sec:simres} 
\begin{figure}[t]
\includegraphics[width=8cm]{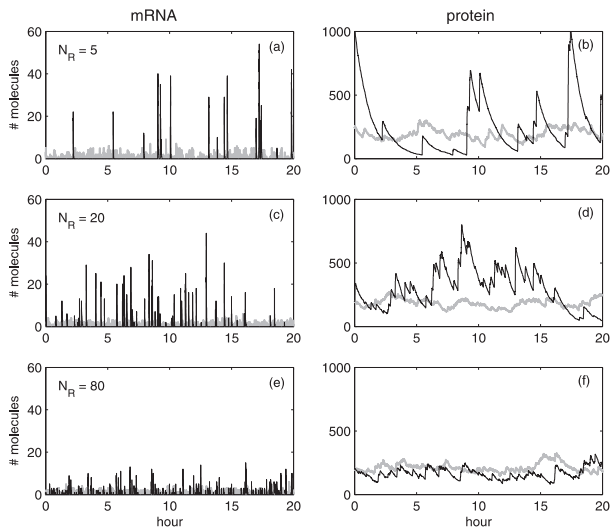} 
\caption{Dynamics of mRNA and protein numbers in the repressed state for different number of 
repressors $N_R$. The number of mRNA and protein molecules is shown for simulations with GFRD (black line) and according to the master 
equation (gray line). In the GFRD simulation, diffusion of repressor particles is explicitly included. (a) and (b) $N_R=5$. (c) and (d) 
$N_R=20$. (e) and (f) $N_R=80$. In general, there is a dramatic difference in dynamics due to the spatial fluctuations of the repressor 
molecules. This difference becomes more pronounced as the number of repressors decreases. However, we find that in all cases $\langle 
N_\text{M} \rangle = 0.5$ and $\langle N_P \rangle = 200$, on average.} \label{fig:tracks}
\end{figure}

To study the effect of spatial fluctuations on the repression of genes, we simulate the reaction network described in Eqs. 
\ref{eq:gene_exp1}-\ref{eq:gene_exp9} both by GFRD, thus explicitly taking into account the diffusive motion of the repressor particles, 
and according to the well-stirred model, where the repressor particles are assumed to be homogeneously distributed in space and the 
dynamics depends only on the concentration of repressor. In Fig. \ref{fig:tracks} we show the behavior of mRNA and protein numbers for 
a system with open complex formation rate $k_\text{OC}=30 s^{-1}$ and with varying numbers of repressors $N_R$. We keep the repression 
factor fixed at $f=100$ so that with increasing $N_R$ the repressor backward rate $k_\text{bR}$ is also increased, {\em i.e.} repressor 
particles are bound to the DNA for a shorter time. 

It is clear from Fig. \ref{fig:tracks} that there is a dramatic 
difference between the behavior of mRNA and protein numbers between 
the GFRD simulation and the well-stirred model. When spatial 
fluctuations of the repressor molecules are included, mRNA is no 
longer produced in a continuous fashion, but instead in sharp, 
discontinuous bursts during which the mRNA level can reach levels 
comparing to those of the unrepressed state. These bursts in mRNA 
production consequently lead to peaks in protein number. As the 
protein decay rate is much lower than that of mRNA, these peaks are 
followed by periods of exponential decay over the course of hours. Due 
to these fluctuations, protein numbers often reach levels of around 
$5-10\%$ of the protein levels in the unrepressed state. In contrast, 
in the absence of repressor diffusion, the fluctuations around the 
average protein number are much lower. For both cases, however, the 
average behavior is identical: even though the dynamics is very 
different, we always find that on average $\langle N_\text{mRNA} 
\rangle = 0.5$ and $\langle N_P \rangle = 200$. Also, in all cases the 
fluctuations in mRNA number are larger than those in protein 
number. This means that the translation step functions as a low-pass 
filter to the repressor signal. 
 
When we increase the number of repressors $N_R$ and change $k_\text{bR}$ in such a way that the repression level $f$ remains constant, 
we find that both for GFRD and the well-stirred model the fluctuations in mRNA and protein number decrease. In the absence of spatial 
fluctuations this effect is minor, but for GFRD this decrease is sharp: for large number of repressors, the burst in mRNA become both 
weaker and more frequent. This in turn leads to smaller peaks and shorter periods of exponential decay in protein numbers. In fact, as 
$N_R$ is increased both approaches converge to the same behavior. At around $N_R \approx 100$, the dynamics of the protein number is 
similar for the well-stirred model and the spatially resolved model. The same happens for mRNA number when $N_R \approx 500$. 
\begin{figure}[t]
\includegraphics[width=8cm]{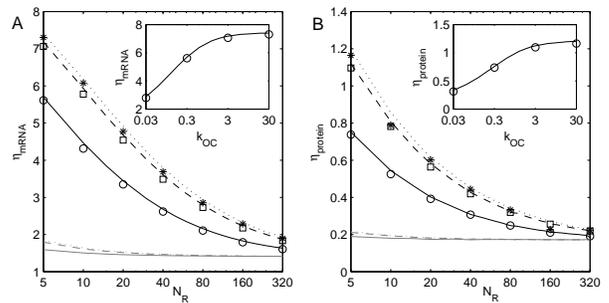} 
\caption{Noise in (a) mRNA number 
and (b) protein number as a function of the number of repressors $N_R$ 
and for constant repression factor $f=100$. Data obtained by GFRD 
simulation is shown for $k_{\rm OC}=0.3 (\circ), 3 (\Box)$ and $30 (*) 
s^{-1}$.  Noise levels for the well-stirred model are shown as grey 
lines and those for the well-stirred model with reaction rates 
renormalized according to Eqs. \ref{eq:renorm1} and \ref{eq:renorm2} 
are shown as black lines, both for $k_{\rm OC} = 0.3$ (solid lines), 
$3$ (dashed lines) and $30$ (dotted lines) $s^{-1}$. Only when the 
reaction rates are properly renormalized does the noise in the 
well-stirred model agree well with the noise in the GFRD simulations, 
which include the effect of diffusion. (Insets) Noise levels as a 
function of $k_{\rm OC}$.  Symbols indicate results for GFRD and lines 
are results for the chemical master equation with renormalized 
reaction rates.} \label{fig:noise} 
\end{figure} 
In Fig. \ref{fig:noise}, we quantify the noise in mRNA and protein 
number, defined as standard deviation divided by the mean, while we 
change the number of repressors $N_R$. As we keep the amount of 
repression fixed at $f=100$, we simultaneously vary the backward rate 
$k_{\rm bR}$ according to Eq.  \ref{eq:repression_factor}. When all 
parameters are the same, the noise for the GFRD simulation, including 
the diffusive motion of the repressors, is always larger than the 
noise for the well-stirred model, where the diffusive motion is 
ignored. In both cases, the noise decreases when the number of 
repressors is increased and the repressor backward rate becomes 
larger. This is consistent with the mRNA and protein tracks shown in 
Fig. \ref{fig:tracks}. We also investigated the effect of changing the 
open complex formation rate $k_{\rm OC}$. In nature, this rate can be 
tuned by changing the base pair composition of the promoter region on 
the DNA. When we change $k_{OC}$, we change the mRNA decay rate 
$k_{\rm dm}$ so that the average mRNA and protein concentrations 
remain unchanged (see section~\ref{sec:TT}). We find that when $k_{\rm 
OC}$ is lowered, the fluctuations in mRNA and protein levels are 
sharply reduced. When $k_{\rm OC}$ is much larger than the RNAP 
backward rate $k_{\rm bRp}=0.5 s^{-1}$, almost every RNAP binding to 
the promoter DNA will result in transcription of a mRNA. For $k_{\rm 
OC}$ smaller than $k_{\rm bRp}$, RNAP binding will lead to 
transcription only infrequently. As a consequence, the operator 
filters out part of the fluctuations in RNAP binding due to the 
diffusive motion of the repressor particles, leading to the decrease 
in noise observed in Fig. \ref{fig:noise}. This shows that the open 
complex formation rate plays a considerable role in controlling noise 
in gene expression. 
\section{Simulations results: operator binding} 
\label{sec:operbind} 
 
\begin{figure}[t]
\includegraphics[width=8cm]{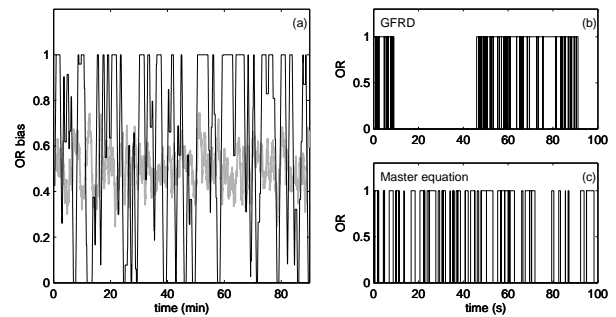} 
\caption{Dynamics of repressor binding for a repression factor of $f=2$ and 
$N_R=5$. (a) The $OR$-bias for GFRD (black line) and the well-stirred 
model (gray line). The $OR$-bias is defined as the fraction of time a 
repressor is bound to the operator site in the last $50$ seconds. When 
the diffusive motion of repressor molecules is included (black line), 
the $OR$-bias switches between periods where repressors are 
continuously bound to or absent from the DNA for long times. (b) and 
(c) Time trace of the occupancy of the operator site by repressor 
molecules. When $OR=1$ a repressor is bound to the operator site and 
$OR=0$ indicates either a free operator site or one with RNAP 
bound. For the GFRD simulations, an initial binding is followed by 
several rapid rebindings, whereas for the well-stirred model binding 
and rebinding is much more unstructured.  Note that here, for reasons 
of clarity, $f=2$ instead of $f=100$ as used in the text and 
Figs. \ref{fig:tracks} and \ref{fig:noise}.} \label{fig:rebinding}
\end{figure}

To understand how the diffusive motion of repressor molecules leads to increased fluctuations in mRNA and protein numbers, it is useful 
to look in some detail at the dynamics of repressor-DNA binding. In figure \ref{fig:rebinding}A, we show the $OR$ bias for both GFRD and 
the well-stirred model. The $OR$ bias is a moving time average over $OR(t)$ with a $50 s$ time window and should be interpreted as the 
fraction of time the operator site was bound by repressor particles over the last $50$ seconds. The results we show here are for $N_R=5$ 
repressors and a repression factor $f=2$. At this repression factor, $k_\text{bR}$ is such that the repressor molecules are bound to 
the operator only fifty percent of the time, making it easier to visualize the operator dynamics than in the case of $f=100$ as used 
above. 
The $OR$ bias for the well-stirred model fluctuates around the average 
value $\langle OR \rangle=0.5$, indicating that on the timescale of 
$50 s$ several binding and unbinding events occur, in agreement with 
$k_\text{bR}=1.26 s^{-1}$ for $f=2$. On the other hand, when 
including spatial fluctuations, the $OR$ bias switches between periods 
in which repressors are bound to the DNA continuously and periods in 
which the repressors are virtually absent, both on timescales much 
longer than the $50 s$ time window. How is it possible that repressors 
are bound to the operator site for times much longer than the 
timescale set by the dissociation rate from the DNA? The answer to 
that question can be found in Figs. \ref{fig:rebinding}B and C, where 
a time trace is shown of the operator occupancy by the repressor for 
both GFRD and the well-stirred model. The time trace for the 
simulation of the well-stirred model in Fig.  \ref{fig:rebinding}C 
shows a familiar picture: binding and dissociation of the repressor 
from the operator occurs irregularly, the time between events given by 
Poisson distributions. The time trace for GFRD in 
Fig. \ref{fig:rebinding}B looks rather different. Here, in general a 
dissociation event is followed by a rebinding very rapidly. Only 
occasionally does a dissociation result in the operator being unbound 
by repressors for a longer time. When this happens, repressors stay 
away from the operator for a time much longer than the typical time 
separating binding events in Fig. \ref{fig:rebinding}C. These series 
of rapid rebindings followed by periods of prolonged absence from the 
operator result in the aberrant $OR$ bias shown in 
Fig. \ref{fig:rebinding}A. 
 
The occurrence of rapid rebindings is intimately related to the nature 
of diffusion. When diffusion and the positions of the reactants are 
ignored all dynamics is based only on the average concentration of the 
reactants. As a consequence, when in this approach a repressor 
dissociates from the operator site, the probability of rebinding 
depends only on the concentration of repressor in the cell.  On the 
level of actual positions of the reactants, this amounts to placing 
the repressor at a random position in the container. The situation is 
very different for the GFRD approach, where the positions of the 
reactants are taken into account. After a dissociation from the 
operator site, the repressor particle is placed at contact with the 
operator site. Because of the close proximity of the repressor to its 
binding site, it has a high probability of rapidly rebinding to, and 
only a small probability of diffusing away from,  the binding site. At 
the same time, when the repressor eventually diffuses away from the 
operator site, the probability that the same, or more likely, another 
repressor diffuses to and binds the operator site is much smaller than 
the probability of binding in the well-stirred model, as will be shown 
quantitatively in Sec. \ref{sec:twostep}. This results in the behavior 
observed in Fig.  \ref{fig:rebinding}B. 
 
It can now be understood that the bursts in mRNA production correspond 
to the prolonged absence of repressor from the operator site compared 
to the well-stirred model. Especially for low repressor 
concentrations, these periods of absence can be long enough that the 
concentration of mRNA reaches values comparable to those in the 
unrepressed state for brief periods of time. When a repressor binds to 
the operator site, due to the rapid rebindings it will remain bound 
effectively for a time much longer than the mRNA lifetime, leading to 
long periods where mRNA is absent in the cell. This shows that under 
these conditions spatial fluctuations and not stochastic chemical 
kinetics are the dominant contribution to the noise in mRNA and 
protein numbers in the repressed state. 
\section{Two-step kinetic scheme} 
\label{sec:twostep} 
 In this section we investigate to what extent the effect of diffusion on the repressor dynamics can be modeled by the two-step kinetic 
scheme \cite{Eigen74, Shoup82}: 
\begin{equation} 
\label{eq:two_step_kinetic_scheme} 
O + R \overset{k_{+}}{\underset{k_{-}}{\rightleftarrows}} O \cdots R \overset{k_{a}}{\underset{k_{d}}{\rightleftarrows}} OR. \\ 
\end{equation} 
The first step in Eq. \ref{eq:two_step_kinetic_scheme} describes the 
diffusion of repressor to the operator site resulting in the encounter 
complex $O \cdots R$, with the rates $k_+$ and $k_-$ depending on the 
diffusion coefficient $D$ and the size of the particles.  The next 
step describes the subsequent binding of repressor to the DNA. In this 
case the rates are related to the microscopic rates defined in 
Eq. \ref{eq:gene_exp1}. When the encounter complex is assumed to be in 
steady state, the two-step kinetic scheme can be mapped onto the 
reaction described in Eq. \ref{eq:gene_exp1}, but with effective rate 
constants $k'_{fR} = k_+ k_{a}/(k_-+k_{a})$ and $k'_{bR} = k_- 
k_{d}/(k_-+k_{a})$ \cite{Eigen74}.  The two-step kinetic scheme should 
yield the same average concentrations as the scheme in 
Eq. \ref{eq:gene_exp1}, so that the equilibrium constant 
$K=k_a/k_d=k'_{fR}/k'_{bR}=k_{fR}/k_{bR}$, where $k_{\rm fR}$ and 
$k_{\rm bR}$ are the reaction rates defined in Eq. \ref{eq:gene_exp1}. 
 
It is possible to express the effective rate constants $k'_{\rm fR}$ 
and $k'_{\rm bR}$ in terms of the microscopic rate constants $k_{\rm 
fR}$ and $k_{\rm bR}$. For the setup used here, where a single 
operator O is surrounded by a homogeneous distribution of repressor R, 
the rate $k_+$ follows from the solution of the steady state diffusion 
equation with a reactive boundary condition with rate $k=k_a$ at 
contact \cite{Smoluchowski17, Shoup82} and is given by the 
diffusion-limited reaction rate $k_D=4 \pi \sigma D$. The rates $k_-$ 
and $k_a$ depend on the exact definition of the encounter complex $O 
\cdots R$. It is natural to identify the rate $k_d$ with the intrinsic 
dissociation rate $k_{bR}$, thus $k_d=k_{bR}$. From these expressions 
for $k_+$ and $k_d$ and the requirement that the equilibrium constant 
should remain unchanged, one finds that $k_a/k_- = k_{fR}/k_D$. Using this 
result one obtains $k'_{fR} = k_D k_{fR}/(k_D + k_{fR})$ and $k'_{bR} 
= k_D k_{bR}/(k_D + k_{fR})$. 
 
These renormalized rate constants have a clear interpretation. For the 
effective forward rate it follows, for 
instance, that: $1/k'_{fR} = 1/k_D + 1/k_{fR}$: that is, on average, 
the time required for repressor binding is given by the time needed to 
diffuse towards the operator plus the time for a reaction to occur 
when the repressor is in contact with the operator site 
\cite{Shoup82}. The effective backward rate has a similar 
interpretation. The probability that after dissociation the repressor 
diffuses away from the operator site and never returns is given by 
$S_{\rm irr}(t \rightarrow \infty| \sigma)$, where $S_{\rm irr}(t, 
r_0)$ is the irreversible survival probability for two reacting 
particles \cite{Kim99}. Using that $S_{\rm irr}(t \rightarrow \infty | 
\sigma) = k_D/(k_{fR}+k_D)$, the expression for $k'_{bR}$ can be 
written as $k'_{bR} = k_{bR}S_{\rm irr}(t \rightarrow \infty| 
\sigma)$: that is, the effective dissociation rate is the microscopic 
dissociation rate multiplied by the probability that after 
dissociation the repressor escapes from the operator site 
\cite{Shoup82}. 
 
For diffusion limited reactions, such as the reaction considered here, we have that $k_{fR} \gg k_D$. Now, the renormalized rate  
constants reduce to: 
\begin{eqnarray} 
\label{eq:renorm1} 
k'_{fR} &=& k_D, \\ 
\label{eq:renorm2} 
k'_{bR} &=& k_D k_{bR} / k_{fR}. 
\end{eqnarray}  
In Fig. \ref{fig:noise}, we compare the noise profiles for the GFRD 
algorithm with those obtained by a simulation of the well-stirred model, 
where instead of the microscopic rates $k_{fR}$ and $k_{fB}$ we use 
the renormalized rates from Eqs. \ref{eq:renorm1} and 
\ref{eq:renorm2}. Surprisingly, we find complete agreement. One of the 
main reasons why this is unexpected, is that for the master equation 
the time between events is Poisson-distributed, whereas after a 
dissociation the time to the next rebinding is distributed according 
to a power-law distribution when diffusion is taken into account 
\cite{Kim99}. The reason that this power-law behavior of rebinding 
times is not of influence on the noise profile, is that the time scale 
of rapid rebinding is much smaller than any of the other relevant time 
scales in the network. Specifically, rebinding times are so short that 
the probability that a RNAP will bind before a rebinding is 
negligible. As a consequence, the transcription network is not at all 
influenced by the brief period the operator site is accessible before 
rebinding: for the transcription machinery the series of consecutive 
rebindings, albeit distributed algebraically in time individually, is 
perceived as a single event. And on much longer time scales, when a 
repressor diffuses in from the bulk towards the operator site, the 
distribution of arrival times is expected to be Poissonian, because on 
these time scales the repressors are distributed homogeneously in the 
bulk. 
 
It is possible to reinterpret the effective rate constants in 
Eq. \ref{eq:renorm1} and \ref{eq:renorm2} in the language of rapid 
rebindings. The probability $p$ that a rebind will occur after a 
dissociation from the DNA is given by $p=1-S_\infty$, where 
$S_t=S_{\rm irr}(t, r_0 = \sigma)$. The probability that $n$ 
consecutive rebindings occur before the repressor diffuses away from 
the operator site is then given by $p_n=(1-S_\infty)^n S_\infty$. From 
this follows that the average number of rebindings is 
$N_{RB}=(1-S_\infty)/S_\infty$. Using again that 
$S_\infty=k_{D}/(k_{fR}+k_D)$, we find that 
$N_{RB}=k_{fR}/k_D$. Combining this with Eqs. \ref{eq:renorm1} and 
\ref{eq:renorm2}, we get: 
\begin{eqnarray} 
\label{eq:renorm3} 
k'_{fR} &=& k_{fR}/N_{RB}, \\ 
\label{eq:renorm4} 
k'_{bR} &=& k_{bR}/N_{RB}. 
\end{eqnarray}  
In words, after an initial binding the repressor spends $N_{RB}$ times 
longer on the DNA than expected on the basis of the microscopic 
backward rate, as it rebinds on average $N_{RB}$ times. Because the 
average occupancy should not change, the forward rate should be 
renormalized in the same way. In conclusion, in this model the effects 
of diffusion can be properly described by a well-stirred model when 
the reaction rates are renormalized by the average number of 
rebindings. 
\section{Power Spectra} 
\label{sec:ps} 
 
In this section, we study how the noise due to the stochastic dynamics 
of the repressor molecules propagates through the different steps of 
gene expression for both the spatially resolved model and the 
well-stirred model. This analysis will also provide further insight 
into why the well-stirred model with renormalized rate constants for 
the (un)binding of the repressor molecules works so well. 
 
In biochemical networks, the noise in the output signal depends upon 
the noise in the biochemical reactions that constitute the network, 
the so-called intrinsic noise, and on the noise in the input signal, 
called extrinsic noise 
\cite{Elowitz02,Pedraza05,Detwiler00,Paulsson04,Shibata05, 
TanaseNicola05}. In our case, the output signal is the protein 
concentration, while the input signal is provided by the repressor 
concentration.  The intrinsic noise arises from the biochemical 
reactions that constitute the transcription and translation 
steps. Moreover, we consider the noise in the protein concentration 
that is due to the (un)binding of the RNAP to (from) the DNA to be 
part of the intrinsic noise. The extrinsic noise is provided by the 
fluctuations in the binding of the repressor to the operator, {\em 
i.e.} in the state OR. Since the total repressor concentration, 
$[R_{\rm T}] = [R] + [OR]$, is constant, the extrinsic noise is also 
given by the fluctuations in the concentration of unbound repressor. 
 
The noise properties of biochemical networks are most clearly 
elucidated via the power spectra of the time traces of the copy 
numbers of the components. Recently, we have shown that if the 
fluctuations in the input signal are uncorrelated with the noise in 
the biochemical reactions that constitute the processing network, 
 the power spectrum of the output signal is given 
by~\cite{TanaseNicola05} 
\begin{equation} 
\label{eq:Sadd} 
S_{\rm P} (\omega) = S_{\rm in} (\omega) + g(\omega) S_{\rm ex} (\omega). 
\end{equation} 
Here, $S_{\rm P}(\omega)$ is the power spectrum of the output signal, 
the protein concentration. The spectrum $S_{\rm in} (\omega)$ denotes 
the intrinsic noise of the processing network; it is defined as the 
noise in the output signal in the absence of noise in the input 
signal. Here, the intrinsic noise is due to the biochemical reactions 
of transcription and translation. The spectrum $S_{\rm ex} (\omega)$ 
is the power spectrum of the input signal, which, in this case, is 
given by the noise in the concentration of unbound repressor: $S_{\rm ex} 
(\omega) = S_{\rm R} (\omega)$; because the total repressor 
concentration is constant this power spectrum is also directly related 
to that of the repressor-bound state of the operator, $S_{\rm OR} 
(\omega)$. The function $g(\omega)$ is a transfer function, which 
indicates how fluctuations in the input signal are transmitted towards 
the output signal. If the extrinsic noise is uncorrelated with the 
intrinsic noise, then $g(\omega)$ is an intrinsic quantity that only 
depends upon properties of the processing network, and not upon 
properties of the incoming signal~\cite{TanaseNicola05}. However, for the network studied 
here, the noise in the input signal is not uncorrelated with the 
intrinsic noise~\cite{TanaseNicola05}. As we have shown recently, this 
means that Eq.~\ref{eq:Sadd} is not strictly 
valid~\cite{TanaseNicola05}; the extrinsic contribution to the power 
spectrum of the output signal can no longer be factorized into a 
function that only depends upon intrinsic properties of the 
network, $g(\omega)$, and one that only depends upon the input signal, 
$S_{\rm ex}(\omega)$. This relation is nevertheless highly 
instructive. Indeed, Eq.~\ref{eq:Sadd} could be interpreted as a 
heuristic definition of the transfer function $g(\omega)$. 
 
The diffusive motion of the repressor molecules impede an analytical 
evaluation of the power spectrum for the extrinsic noise. Moreover, 
while power spectra can be calculated analytically for linear reaction 
networks~\cite{Warren05_2}, the delays in transcription resulting from 
promoter clearance and elongation, preclude the derivation of an 
analytical expression for the power spectrum of the intrinsic 
noise. We have therefore obtained the power spectra $S_{\rm P} (\omega)$, 
$S_{\rm ex} (\omega)$, and $S_{\rm in} (\omega)$, directly from the time traces 
of the copy numbers. The power spectrum of a component ${\rm X}$ is given by $S_{\rm X} 
(\omega)=\langle |\tilde X(\omega)|^2\rangle$, where $\tilde 
X(\omega)$ is the Fourier Transform of the concentration $X(t)$ of 
component ${\rm X}$. Conventional FFT algorithms are not convenient, 
because our signals vary over a wide range of time scales. We 
therefore adopted a novel and efficient procedure, which is 
described in the appendix. This procedure should prove useful for 
computing the power spectra of time traces of copy numbers of species in 
biochemical networks, as obtained by Kinetic Monte Carlo simulations.

As indicated above, the intrinsic noise, $S_{\rm in} (\omega)$, is 
defined as the noise in the output signal in the absence of 
fluctuations in the input signal. In order to determine the intrinsic 
contribution to the noise in the protein concentration, we discarded 
the (un)binding reaction of the repressor to the DNA 
(Eq. \ref{eq:gene_exp1}), while rescaling the rate $k_{\rm bRp}$ for 
the dissociation reaction of the RNAP from the DNA 
(Eq. \ref{eq:gene_exp2}) in such a way that the average concentration 
of the protein $P$ remains unchanged. This eliminates the extrinsic noise arising 
from the repressor dynamics, thereby allowing us to obtain the 
intrinsic noise of the reactions in 
Eqs. \ref{eq:gene_exp2}-\ref{eq:gene_exp9}. The rescaled backward rate 
$k^*_{\rm bRp}$ is given by 
\begin{equation} 
k^*_{\rm bRp} = k_{\rm bRp}(1+K_2N_R/V)+k_{\text{OC}}K_2/V 
\end{equation} 
where $K_2=k_{\rm fR}/k_{\rm bR}$.

For the interpretation of the power spectra of the mRNA and protein 
concentration, as discussed below, it is 
instructive to recall the power spectrum of a linear birth-and-death 
process, 
\begin{equation} 
\label{eq:bdp} 
\varnothing \overset{k}{\rightarrow} A \overset{\mu}{\rightarrow} 
\varnothing, 
\end{equation} 
with rate constants $k$ and $\mu$. For the interpretation of the 
spectra of repressor binding to the DNA, it is useful to recall the 
spectrum of a two-state model, 
\begin{equation} 
\label{eq:2ss} 
O \overset{k_1}{\underset{k_2}{\rightleftarrows}} O^*, 
\end{equation} 
with rate constants $k_1$ and $k_2$. For both models, the power 
spectrum  is a Lorentzian 
function of the form: 
\begin{equation} 
\label{eq:Sbd2s} 
S(\omega)=\frac{2 \sigma^2 \mu}{\mu^2+\omega^2}. 
\end{equation} 
 For the birth-and-death process, the variance in the 
concentration of $A$, $\sigma^2$, is $k/\mu$, while for the two state 
system, the variance $\sigma^2$ in the occupancy $n$ is $n(1-n)$; the 
decay rate in the two-state model is $\mu = k_1 + k_2$. The corner 
frequency $\mu$ yields the time scale on which fluctuations relax back 
to steady state. We also note that the noise strength $\sigma^2$ is 
given by the integral of the power spectrum $S (\omega)$: $\sigma^2 = 
1/(2 \pi) \int_{-\infty}^{\infty} d\omega S (\omega)$. The noise 
strength is thus dominated by those frequencies at which the power 
spectrum is largest.  
 
In the next subsection, we discuss the effect of spatial fluctuations 
on the noise in gene expression and explain why a well-stirred model 
with renormalized rate constants for repressor (un)binding can capture 
 its effect. In the subsequent section, we discuss how the noise is 
propagated through the different stages of gene expression. 
\subsection{Spatial Fluctuations} 
\begin{figure}[t]
\includegraphics[width=8cm]{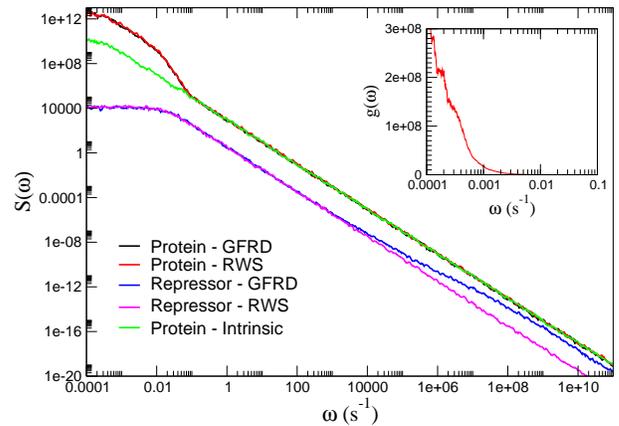} 
\caption{Power spectra of the repressor and protein concentrations obtained for 
$f=100$, $k_{\text{OC}}=30 s^{-1}$, $N_R=5$. Data are shown both for 
the renormalized well-stirred model (RWS) with reaction rates 
renormalized according to Eqs. \ref{eq:renorm1}-\ref{eq:renorm2}, and 
for GFRD, taking into account the spatial fluctuations of the 
repressor molecules explicitly. Also shown is the power spectrum of 
the intrinsic noise, which is the power spectrum of the protein 
concentration in the absence of fluctuations in the repressor 
concentration (extrinsic noise). For large $\omega$, the repressor 
spectrum (extrinsic noise) differs between the well-stirred and the 
spatially resolved model. However, this difference does not appear in 
the power spectra of the protein concentration. The inset shows the 
frequency-dependent gain $g(\omega)$ (see Eq.~\ref{eq:Sadd}).}\label{fig:PS1} 
\end{figure}

 In Fig. \ref{fig:PS1}, we show the power spectra for the input and 
output signals, for both the spatially resolved model and the 
well-stirred model with renormalized rate constants for repressor 
(un)binding (see previous section). We recall that the output signal 
is the protein concentration, while the input signal is the 
concentration of unbound repressor (the extrinsic 
noise). Fig.~\ref{fig:PS1} also shows 
the power spectrum of the intrinsic noise. This is the noise in the 
protein concentration (the output signal), when the noise in the input 
signal resulting from the repressor dynamics has been eliminated by 
the procedure outlined above. The power spectra have been obtained in 
a parameter regime where the diffusing repressors have a large effect 
on the noise: $k_{\text{OC}}=30 s^{-1}$, $N_R=5$ (see 
Fig.~\ref{fig:noise}). 
 
Fig.~\ref{fig:PS1} shows that the power spectrum of the protein 
concentration in the spatially resolved model is identical to that in the 
well-stirred model for the entire range of frequencies observed. This 
confirms the observation in Section \ref{sec:twostep} that the effect 
of the spatial fluctuations of the repressor molecules on the noise in 
the protein concentration can by described by a well-stirred model in 
which the reaction rates for repressor (un)binding to the DNA are 
properly renormalized. 
 
Fig.~\ref{fig:PS1} also elucidates the reason why a well-stirred model 
with properly renormalized rate constants for repressor (un)binding 
can successfully describe the effect of the diffusive motion of the 
repressor molecules on the noise in gene expression.  It is seen that 
the repressor spectrum for the renormalized well-stirred model is 
accurately described by a Lorentzian function with a corner frequency 
$\mu=0.02 s^{-1}$, as expected for the dynamics of repressor 
(un)binding dynamics (see next section). The repressor spectrum of the 
spatially resolved model fully overlaps with that of the well-stirred 
model up to a frequency of $\omega \approx 10^{6} {\rm s}^{-1}$, but for 
higher frequencies it shows a clear deviation from the $\omega^{-2}$ 
behavior.  This deviation is caused by the diffusive motion of the 
repressor molecules. Indeed, the deviation occurs at frequencies 
comparable to the inverse of the typical time scale for rapid 
rebindings ($\sim \mu{\rm s}$). However, this difference between the 
spectrum of the repressor dynamics in the spatially resolved model and 
that in the well-stirred model does not manifest itself in the spectra 
for the protein concentrations of the two respective models, for two 
reasons: 1) the difference only occurs at high frequencies, {\em i.e.} 
in a frequency regime where the fluctuations only marginally 
contribute to the noise strength (the difference in area under the 
curves of the repressor power spectra for the two models is less than 
$5 \%$); 2) the repressor fluctuations in this frequency range are 
filtered out by the processing network of transcription and 
translation; as a result of this, the effect of the small difference 
in area under the curves of the repressor power spectra for the two 
models is reduced even further.  The filtering properties of the 
processing network are illustrated in the inset of Fig.~\ref{fig:PS1}, 
which shows the transfer function $g(\omega)$ as obtained from 
$g(\omega)=(S_P(\omega)-S_{\text{in}}(\omega))/S_{\text{ex}}(\omega)$ 
(see Eq.~\ref{eq:Sadd}). Clearly, the transfer function rapidly 
decreases as the frequency increases. This shows that the processing 
network of transcription and translation acts as a low-pass filter, 
rejecting the high frequency noise in the repressor dynamics that 
originates from the rapid rebindings. 
\begin{figure}[t]
\includegraphics[width=8cm]{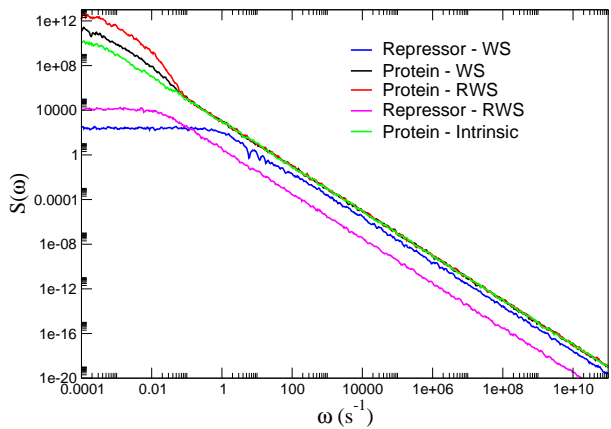}
\caption{The power spectra for the well-stirred model with unrenormalized rate 
constants (WS) and for the well-stirred model with renormalized rate 
constants for repressor (un)binding to (from) the DNA (RWS). The 
intrinsic noise of gene expression is the same for both models. The 
extrinsic noise, arising from the repressor dynamics, is, however, 
markedly different. The repressor spectrum for the well-stirred model 
with renormalized rate constants has lower corner frequency, but, more 
importantly, also a higher power at low frequencies. The increased 
power at low frequencies is not filtered by the processing network and 
increases the noise in gene expression.} 
\label{fig:PS3} 
\end{figure} 
 The only effect of the repressor rebindings on the noise in gene 
expression is thus that it lowers the effective dissociation rate (and 
association rate), as explained in the previous section. As compared 
to the well-stirred model with the unrenormalized rate constants for 
repressor (un)binding, this decreases the corner frequency $\mu$ in 
the repressor power spectrum (see Fig.~\ref{fig:PS3}), but {\em 
increases} the power at low frequencies -- recall that for a two-state 
model, which relaxes mono-exponentially, the power spectrum at zero 
frequency is $S (\omega = 0) = 2 \sigma^2/\mu$, which thus increases 
as the relaxation rate $\mu = k_1 + k_2$ decreases as a result of the 
slower binding and unbinding of repressor (see Eq.~\ref{eq:Sbd2s}). The higher 
power in the repressor spectrum at low frequencies for the spatially 
resolved model and for the well-stirred model with the renormalized 
rate constants, as compared to that for the well-stirred model with 
the unrenormalized rate constants, is not filtered by the processing 
network of transcription and translation and thus manifests itself in 
the power spectrum of the protein concentration. Spatial fluctuations 
of gene regulatory proteins thus increase the noise in gene expression 
by increasing the power of the input signal at low frequencies. 
\subsection{Noise propagation} 
\begin{figure}[t]
\includegraphics[width=8cm]{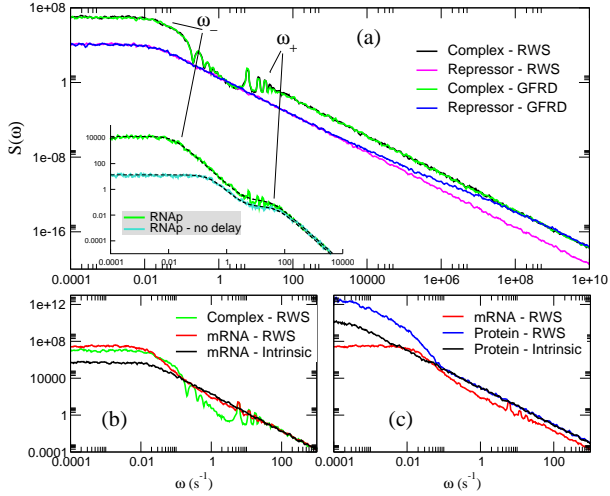} 
\caption{Comparison of the power spectra at different stages of gene 
expression. (a) Power spectrum for repressor concentration and for the 
elongation complex $ORp^*+T$, both for the well-stirred model with 
renormalized rate constants (RWS) and for GFRD.  Repressor power 
spectra show a difference between the spatially resolved model and the 
well-stirred model at high frequencies, due to the diffusion of the 
repressor molecules. The power spectra for the elongation complexes 
coincide for the well-stirred and the spatially resolved model. The 
power spectrum of the elongation complex shows a series of peaks and 
valleys due to the presence of fixed delays in the dynamics of the 
elongation complex. (inset) Power spectrum of RNAP dynamics ($OR_p + 
OR_p^*$). Shown are the power spectra in the presence and absence of 
fixed delays in the RNAP dynamics. Due to the competition between RNAP 
and rerepssor for binding to the promoter, the power spectrum is 
described by a sum of two Lorentzians. (b) Power spectra of the 
elongation complex and mRNA. Peaks due to the delays in RNAP dynamics 
are still present in the mRNA dynamics. For high frequencies, the mRNA 
dynamics is well described by a linear birth-and-death process. (c) 
Spectra of mRNA and protein. The slow protein dynamics filters out all 
the peaks resulting from the delays in the RNAP dynamics. The only 
difference between the full spectrum of the output signal and that of 
the intrinsic noise is an increased noise at low frequencies, 
due to the repressor dynamics.}\label{fig:PS2} 
\end{figure}

In Fig. \ref{fig:PS2} we show how fluctuations in the input signal 
arising from the dynamics of repressor binding and unbinding, are 
propagated through the different stages of gene expression. In 
Fig. \ref{fig:PS2}(a) we illustrate how the noise in the repressor 
concentration (the extrinsic noise) is transferred to the level of 
transcription. The figure shows for both the spatially resolved model 
and for the well-stirred model with renormalized rate constants for 
repressor (un)binding, the power spectrum of the repressor 
concentration and the spectrum of the concentration of the elongation 
complex, defined as $[ORp^*]+[T]$.  It is clear from 
Fig. \ref{fig:PS2}(a) that already at the level of the elongation 
complex, the high-frequency noise due to the rapid rebindings is 
filtered. Transcription can thus already be described by a 
well-stirred model with properly renormalized rate constants for 
repressor (un)binding to (from) the DNA. 
The power spectrum of the elongation complex exhibits two corner 
frequencies, one around $\omega_+ \approx 40 s^{-1}$ and another one 
at $\omega_- \approx 0.02 s^{-1}$. These two corner frequencies arise 
from the competition between repressor and RNAP for binding to the 
promoter. To elucidate this, we have plotted in the inset the power 
spectrum for RNAP bound to the promoter, thus the power spectrum for 
$[ORp] + [ORp^*]$. It is seen that this power spectrum has the same 
two corner frequencies as that of the elongation complex, showing that 
their dynamics is dominated by the same processes -- repressor binding 
and RNAP binding to the promoter. These two corner frequencies can be 
estimated analytically by considering the reactions in 
Eqs.~\ref{eq:gene_exp1}-\ref{eq:gene_exp4} as a three-state system, in 
which repressor and RNAP compete for binding to the promoter: 
\begin{equation} 
\label{eq:three_state} 
OR \overset{k_{2}}{\underset{k_{1}}{\rightleftarrows}} O  \overset{k_{3}}{\underset{k_{4}}{\rightleftarrows}} ORp^\prime.  
\end{equation}  
Here, $ORp^\prime = ORp + ORp^*$, where $ORp$ denotes the RNAP bound 
to the promoter in the closed complex and $ORp^*$ denotes RNAP bound 
to the promoter in the open complex. The rate constant $k_1$ denotes 
the rate at which a repressor binds to the promoter; it is given by 
$k_1=k'_{fR}[R_{\rm T}]$, where $k'_{fR}$ is the renormalized 
association rate (see Eq.~\ref{eq:renorm1}). The rate constant $k_2$ 
denotes the renormalized rate for repressor unbinding, $k_2=k'_{bR}$ 
(see Eq.~\ref{eq:renorm2}); $k_3=k_{fRp}$ denotes the rate at which 
RNAP binds to the promoter. The rate constant $k_4$ is the rate at 
which the RNAP leaves the promoter. Since the promoter can become 
accessible for the binding of another RNAP or repressor by either the 
dissociation of RNAP from the closed complex or by forming the open 
complex and then clearing the promoter, this rate is given by 
$k_4=k_{bRp}+(k_{OC}^{-1}+t_{\rm clear})^{-1}$. If promoter clearance 
would be neglected, then, indeed, $k_4=k_{bRp}+k_{OC}$. 
 
The power spectrum of the RNAP dynamics in Eq. \ref{eq:three_state} 
can be calculated analytically and is given by a sum of two 
Lorentzians: 
\begin{equation} 
\label{eq:SORp} 
S_{ORp^{\prime}}(\omega) = \frac{A \omega_-}{\omega_-^2 + \omega^2} + \frac{B \omega_+}{\omega_+^2 + \omega^2}, 
\end{equation} 
where $A$ and $B$ are coefficients. The corner frequencies $\omega_-$ 
and $\omega_+$ are given by $\omega_\pm = (k \pm \sqrt{k^2-4h})/2$, 
where $k=\sum_i k_i$ and $h=k_1 k_4 + k_2 ( k_3 + k_4 )$. The dynamics 
of repressor binding and unbinding is much slower than that of RNAP 
binding and unbinding, meaning 
that $k_1,k_2 \ll k_3, k_4$. This allows us to approximate the corner 
frequencies as $\omega_+ = k_3 + k_4$ and $\omega_- = k_2 + k_1 k_4 / (k_3 + 
k_4)$. This yields the  following expressions for the corner frequencies: 
\begin{eqnarray} 
\omega_+ &=& k_{fRp} + k_{bRp} + (k_{OC}^{-1}+t_{\rm clear})^{-1}\\ 
\omega_- &=& k_{bR}^\prime + k^\prime_{fR} [R_{\rm T}] [O]^{\prime}. 
\end{eqnarray} 
Here, $[O]^\prime \equiv k_4 / (k_3 + k_4)$ is the conditional 
probability that the promoter is not occupied by the RNAP, given that 
it is not occupied by repressor; it is given by the occupancy of the 
promoter by RNAP in the {\em absence} of any repressor molecules in 
the system.  We can now see that the highest corner frequency, 
$\omega_+$, describes the {\em fast} dynamics of RNAP binding to, and 
clearing from, the promoter and that the other corner frequency, 
$\omega_-$, represents the {\em slow} dynamics of repressor 
(un)binding to the DNA in the {\em presence} of the fast RNAP bindings 
to the promoter; the lower corner frequency, $\omega_-$, is also the 
corner frequency in the repressor spectrum of the renormalized 
well-stirred model (see Figs.~\ref{fig:PS1} and~\ref{fig:PS3}). In 
Fig.~\ref{fig:PS2}(a) we plot the power spectrum $S_{ORp^\prime} 
(\omega)$ as predicted by the three-state model (Eq.~\ref{eq:SORp}; 
with fitted coefficients $A$ and $B$) on top of the power spectrum 
obtained from the simulations and find excellent agreement. We also 
show the power spectra when we neglect the delay due to promoter 
clearance. As expected, in the absence of the delay due to promoter 
clearance, the lower corner frequency, $\omega_-$, and, to a smaller 
extent, the higher corner frequency, $\omega_+$, are shifted to higher 
frequencies. 
 
The power spectrum of the elongation complex in Fig. \ref{fig:PS2}(a) 
contains information that is not easily observed in the time domain 
and could as a result be helpful in the interpretation of the 
results. It is seen that there are two series of peaks. Those are 
associated with the two processes with fixed time delays. The first 
process is the promoter clearance, which takes a fixed time $t_{\rm 
clear}$. Indeed, the first peak in the corresponding series of peaks 
in the power spectrum of the elongation complex, is at $\omega\approx 
2\pi/(t_{\text{clear}})=6.3 s^{-1}$; the other peaks in the series are 
the higher harmonics that naturally arise for processes with fixed 
time delays.  The second process is the transcript elongation 
process. After the elongation complex has been formed, it takes a 
fixed time $t_{\rm clear} + t_{\rm elon}$ before the full transcript 
is formed and the RNAP dissociates from the DNA; the first valley of 
the corresponding series of peaks/valleys is, indeed, at $\omega \approx 
2 \pi /(t_{\rm clear} + t_{\rm elon}) = 0.2 {\rm s}^{-1}$. While the 
frequency $2 \pi / t_{\rm clear}$ yields, to a good approximation, the 
rate at which the elongation complex signal increases, the frequency 
$2 \pi /(t_{\rm clear} + t_{\rm elon})$ corresponds to the frequency 
at which the elongation complex signal {\em decreases}; this explains 
why the shapes of the respective series of peaks and valleys are 
reciprocal. Lastly, the reason that both peaks and valleys are 
broadened is that the delay in the formation of the elongation complex 
is not fully deterministic: the duration of the delay is not only 
determined by the promoter clearance time, which, indeed, is fixed, 
but also by the time it takes for another RNAP to bind the DNA and 
then form the open complex -- in the absence of repressor, the average 
frequency at which an elongation complex is formed is given by $2 \pi 
/ (k_{\rm fRp}^{-1} + k_{\rm OC}^{-1} + t_{\rm clear})$ (see also 
Eqs.~\ref{eq:gene_exp2}--~\ref{eq:gene_exp4}). Both RNAP binding and 
open complex formation are modeled as Poisson processes, and this 
leads to a {\em distribution} of delay times for the formation of the 
elongation complex. 
 
In Fig.~\ref{fig:PS2}(b) and (c), we examine how the noise in the 
dynamics of the elongation complex propagates to the level of mRNA and 
protein dynamics. In Fig.~\ref{fig:PS2}(b), we compare the full power 
spectrum of the mRNA concentration with that of the elongation complex 
-- the input signal (extrinsic noise) for the mRNA signal -- and that 
of the intrinsic noise of the mRNA signal; to compute the intrinsic 
noise, we have modeled the mRNA dynamics as a birth-and-death process 
(see Eq.~\ref{eq:bdp}) with a production rate as given by the average 
production rate for the full system in 
Eqs. \ref{eq:gene_exp1}-\ref{eq:gene_exp9}. As expected, for higher 
frequencies ($\omega > 0.1 {\rm s}^{-1}$), the full spectrum of mRNA 
overlaps almost fully with that of the intrinsic noise, although some 
traces of the input signal (the elongation complex) are still apparent 
in this high frequency regime; these are the peaks at $\omega \approx 
6.3 {\rm s}^{-1}$ corresponding to promoter clearance. At lower 
frequencies ($\omega < 0.1 {\rm s}^{-1}$), the noise in the mRNA signal 
is dominated by the extrinsic noise, which is the noise in the 
elongation complex (the input signal). Indeed, both the spectrum of 
the elongation complex and that of mRNA have a corner frequency at 
$\omega_-$, which, as discussed above, arises from the slow repressor 
(un)binding to the DNA in the presence of the fast DNA-(un)binding 
kinetics of RNAP. 
 
Fig.~\ref{fig:PS2}(c) shows how the noise in the mRNA concentration is 
propagated to that in the protein concentration. Again, at higher 
frequencies, the spectrum of the protein concentration coincides with 
that of the intrinsic noise of protein synthesis, which, as above for 
mRNA, has been computed by modeling protein production as a 
birth-and-death process; note also that the remnants of operator clearance 
(the peaks in the spectrum at $\omega \approx 6.3 {\rm s}^{-1}$) have 
been filtered by the slow protein dynamics. Only for frequencies 
smaller than $\omega \approx 0.1 {\rm s}^{-1}$, does the extrinsic 
noise -- the noise in the mRNA concentration -- strongly contribute to 
the noise in the protein concentration. A careful inspection of the 
protein spectrum shows that it has a ``corner'' at $\omega_-$, which 
arises from the repressor DNA-(un)binding dynamics (the extrinsic 
noise), and one, albeit much less visible, at $\omega \approx k_{\rm 
dp} = 2 \times 10^{-4} {\rm s}^{-1}$, which is due to the intrinsic 
dynamics of protein degradation.\section{Discussion and Outlook} 
Our analysis reveals that at high frequencies both mRNA and protein 
synthesis are well described by a linear birth-and-death model. In this 
frequency regime, the effect of spatial fluctuations, originating from 
the rapid repressor rebindings, is completely filtered by the slow 
dynamics of transcription and translation. These rebindings do, 
however, decrease the effective rate at which the repressor molecules 
associate with, and dissociate from, the promoter. This increases the 
intensity of the extrinsic (repressor) noise in the low frequency 
regime. Moreover, the low-frequency fluctuations in the repressor 
binding do propagate through the different stages of gene 
expression. In particular, they lead to sharp bursts in the production of 
mRNA and protein. These bursts increase the noise intensity at the 
lower frequencies in the noise spectrum of mRNA and protein. And since 
the noise strength $\sigma^2$ is dominated by fluctuations in the 
low-frequency regime, spatial fluctuations ultimately strongly 
increase the noise in mRNA and protein concentration. 
 
Recently, experiments have been performed, in which the synthesis of 
individual mRNA transcripts~\cite{Golding05} and individual protein 
molecules~\cite{Yu06} could be detected. The systems in these studies 
were very similar to that studied here: a gene under the control of a 
(Lac) repressor. These studies unambiguously demonstrated that 
transcription~\cite{Golding05} and translation can occur in 
bursts~\cite{Yu06}. Our simulation results show that spatial 
fluctuations of the repressor molecules might be responsible for 
this. Indeed, our results strongly suggest that spatial fluctuations 
are the dominant source of noise in gene expression in these 
systems. 
 
The spatial fluctuations due to diffusion of the repressor molecules 
could have significant implications for the functioning of gene 
regulatory networks. Under some conditions, it might be crucial that 
the protein number is not only low on average, but remains low at all 
times. For instance, if the protein itself functions as a 
transcription factor, it might by accident induce the expression of 
another gene, when, due to a fluctuation, its concentration crosses a 
particular activation threshold. Thus, not all combinations of repressor copy 
number $N_R$ and repressor backward rate $k_\text{bR}$ that obey 
Eq. \ref{eq:repression_factor} and thus have the same average 
repression strength, are necessarily equivalent in terms of function 
when diffusion is taken into account. If the fluctuations in the 
repressed state need to be small, then the cell could increase the 
number of repressors and decrease the binding affinity to the operator 
site, such that the repressor molecules stay bound to the DNA only 
briefly. Alternatively, the cell could minimize the effect of 
fluctuations by reducing the rate at which the open complex is formed 
by RNAP -- our analysis shows that the process of open complex 
formation can act as a strong low-pass filter. 
 
The rapid rebindings observed in our simulations are a general 
phenomenon. We now address the question when the effect of 
spatial fluctuations due to diffusion can be described by a 
well-stirred model in which the association and dissociation rates are 
renormalized. In the current problem, the rebinding time for a 
dissociated repressor is exceedingly short. As a consequence, the 
probability that a RNAP binds to the promoter during this time, is 
vanishingly small. This is precisely the reason that the effective dissociation rate is 
simply the bare dissociation rate divided by the number of rebindings 
(see Eq.~\ref{eq:renorm3}); the effective association rate is 
renormalized accordingly, because the equilibrium constant should 
remain unchanged (see Eq.~\ref{eq:renorm4}). The success of the 
renormalized well-stirred model is thus a result of the strong separation of 
time scales -- the time scale of repressor rebinding is well separated 
from that of RNAP binding. 
 
The separation of time scales also makes it possible to account for 
the effect of spatial fluctuations by renormalizing the association 
and dissociation rates in other cases. For instance, we have simulated 
a system in which repression occurs in a cooperative manner. In this 
system, the repressor backward rate is smaller when two repressors are 
bound to the operator than when a single repressor is bound. However, 
when one of the two repressors dissociates, its rebinding time is so 
short that the probability for the other repressor to dissociate in 
the mean time, is negligible for reasonable values of 
cooperativity. As a result, the effect of spatial fluctuations can be 
described by a well-stirred model with properly renormalized reaction 
rates. We have also studied a system in which the expression of a gene 
is not under the control of a repressor, but rather under the control 
of an activator. Also in this system, diffusion of the 
transcription factors leads to an enhancement of noise in gene 
expression through a similar mechanism. 
 
Do these observations imply that the effect of spatial fluctuations 
can always be described by a well-stirred model?  In the system 
studied here, the ligand (repressor) molecules bind to a single site. 
We expect that the effect of spatial fluctuations becomes more 
intricate when the number of binding sites for a particular ligand 
increases -- the binding of the ligand to the different sites will 
then exhibit correlations. This could be important when the ligand 
binds to receptors that occur in dense clusters, as in bacterial 
chemotaxis~\cite{bray98,Andrews05} and in the immune 
response~\cite{Valitutti95}. In gene regulatory networks this effect 
could also be significant. Recently, we have shown that in {\em 
E. coli}, pairs of co-regulated genes -- genes that are controlled by 
a common transcription factor -- tend to lie exceedingly close to each 
other on the genome~\cite{Warren04_2}: their promoter regions are 
often separated by a distance smaller than a few hundred base 
pairs. It is conceivable that spatial fluctuations of the 
transcription factors introduce correlations between the noise in the 
expression of these pairs of co-regulated genes. This study also 
revealed that pairs of genes that regulate each other, often lie close 
together, again suggesting that the diffusive motion of transcription 
factors could be important for the functioning of gene regulatory 
networks~\cite{Warren04_2}. 
 
Even in the case of a single gene, the effect of spatial fluctuations 
is expected to be more subtle than that reported here. In this study, 
the operator is modeled as a spherical site. However, as mentioned in 
section~\ref{sec:diffusion}, transcription factors are believed to 
find their operator site via a combination of free 3D diffusion and 1D 
sliding along the DNA. While on length scales larger than the sliding 
distance this process is indeed essentially 3D diffusion, on length 
and time scales smaller than the sliding distance and sliding time, 
respectively, the dynamics is more complicated. We expect that sliding 
could have two important effects. First, it will increase the {\em 
number} of rebindings -- the probability that in 1D a random walker 
returns to the origin is one, while in 3D there is a finite 
probability that it will escape and never return. Secondly, sliding is 
expected to also increase the {\em duration} of the rebindings, 
especially when diffusion along the DNA is much slower than diffusion 
in the cytoplasm.  It is thus conceivable that with sliding, the 
non-exponential relaxation of the operator state, arising from the 
rebindings, shifts to lower frequencies (see Fig.~\ref{fig:PS1}), 
where fluctuations in the operator state are not filtered out. In 
addition, it is possible that with sliding, RNAP and repressor compete 
for binding to the promoter on similar time scales, which would mean 
that the effective dissociation rate is no longer simply given by the 
bare dissociation rate divided by the number of rebindings. Indeed, 
under these conditions, the effect of spatial fluctuations is likely 
to become non-trivial, and describing it would probably require a 
spatially resolved model. We leave this for future work. 
 
Finally, we address the question whether spatial fluctuations, and, 
more in particular, the rebindings, could be studied 
experimentally. Interestingly, recent biochemical data on the 
restriction enzyme EcoRV suggests that after an initial dissociation, 
10-100 rebindings occur before the enzyme escapes into the bulk 
solution~\cite{Stanford00,Gowers05}, in good agreement with the 
average number of rebindings calculated in 
section~\ref{sec:twostep}. However, in our gene expression model, the 
rebinding times are so short that it would seem difficult to probe the 
repressor rebindings directly in an experiment. In fact, reaction 
rates measured biochemically will probably already be corrected for 
according to Eqs.~\ref{eq:renorm1} and~\ref{eq:renorm2}. Sliding along 
the DNA, however, may extend the rebinding times to accessible 
experimental time scales. Moreover, recent experiments suggest that 
the motion of proteins in the nucleoid might be sub-diffusive, which 
would increase the importance of the 
rebindings~\cite{Golding06}. Recently, magnetic tweezer experiments on 
a mechanically stretched, supercoiled, single DNA have made it 
possible to study the kinetics of the open complex formation and 
promoter clearance~\cite{Revyakin04}. Performing these experiments on 
a promoter that is under the control of a repressor, seems a promising 
approach for studying the effect of spatial fluctuations due to the 
diffusive motion of transcription factors on the dynamics of gene 
expression. 
\section*{Acknowledgments} We would like to thank Mans Ehrenberg and Frank Poelwijk for useful discussions and a critical reading 
of the manuscript. The work is part of the research program of the 
``Stichting voor Fundamenteel Onderzoek der Materie (FOM)", which is 
financially supported by the ``Nederlandse organisatie voor 
Wetenschappelijk Onderzoek (NWO)".
\section*{Appendix Computing Power Spectra} 
The power spectrum of the time trace of the copy number $X(t)$ of a species 
$\rm X$ can be efficiently computed by exploiting the fact that in 
between the times $t_k$ the signal $X(t)$ is constant. The Fourier Transform 
 $S_{\rm X}(\omega)$ of $X(t)$ is 
\begin{equation} 
\tilde{X}(\omega)=\int X(t) e^{-i\omega t} dt=\sum_k\int_{t_{k-1}}^{t_k}X_k e^{-i\omega t} dt. 
\end{equation} 
As $X(t)$ is constant within every interval $\{t_{k-1},t_k\}$, the 
integration can easily be performed: 
\begin{equation} 
\tilde{X}(\omega)=\sum_k X_k\frac{1}{-i\omega} (e^{-i\omega t_k} -e^{-i\omega t_{k-1}}). 
\end{equation} 
Shifting up by one the index $j$ in the second part of the sum, we obtain: 
\begin{equation} 
\tilde{X}(\omega)=\frac{1}{i\omega}\sum_k (X_{k+1}-X_k)(e^{-i\omega t_k}). 
\end{equation} 
The real and imaginary parts of the Fourier Transform are thus: 
\begin{eqnarray} 
\Re{[\tilde{X}(\omega)]} &=&\frac{1}{\omega}\sum_k \delta_k(\sin{\omega 
t_k})\\ 
\Im{[\tilde{X}(\omega)]}&=&\frac{1}{\omega}\sum_k \delta_k(\cos{\omega t_k}), 
\end{eqnarray} 
where we have defined $\delta_k=X_{k+1}-X_k$. 
The Power spectrum $S_{\rm X} 
(\omega)=\Re{[\tilde{X}(\omega)]}^2+\Im{[\tilde{X}(\omega)]}^2$ is 
thus given by 
\begin{equation} 
\label{eq:PS} 
S_{\rm X} (\omega)=\left(\frac{1}{\omega}\sum_k\delta_k\cos(\omega 
 t_k) \right)^2 + \left(\frac{1}{\omega}\sum_k\delta_k\sin(\omega t_k) 
 \right)^2. 
\end{equation} 
 
The Fourier Transforms were computed at 10000 
logarithmically spaced angular frequencies starting from 
$\omega_{\text{min}}=10\cdot2\pi/T$, where $T$ is the total length of 
the signal. Power spectra obtained according to Eq.~\ref{eq:PS} were 
filtered with a box average over 20 neighboring points. 
 
\bibliographystyle{apsrev}

\end{document}